# The generality of uncooperative and cooperative effects in elementary hydrogen-bonded systems

Rui Liu[1], Rui Wang[1], Danhui Li[1], Yu Zhu[1], Xinrui Yang[1] and Zhigang Wang[1,2,*]

[1]Institute of Atomic and Molecular Physics, Jilin University, Changchun 130012, China
[2]Institute of Theoretical Chemistry, Jilin University, Changchun 130023, China.

*email: wangzg@jlu.edu.cn (Z. W.)

## Abstract

The cooperative effect plays a significant role in understanding the intermolecular donor-acceptor interactions of hydrogen bonds (H-bonds, D-H···A). Herein, using the benchmark method of high-precision ab initio, the well-known cooperative effect is reproduced in elementary H-bonded systems with different D and A atoms. That is, with the decreasing of intermolecular distance, the D-H bond length first increases and then decreases, while the H···A bond length decreases. On the contrary, when D and A are the same, as the intermolecular distance decreases, the D-H bond length decreases without increasing, which is referred to as the uncooperative effect. Further analyses conclude that compared to cooperative H-bonded systems, uncooperative systems at their respective equilibrium position have a larger core-valence bifurcation (CVB) index (>0.022) and lower binding energies (<0.25 eV), showing a clear linear inverse relationship related to H-bond strength. Therefore, the intermolecular non-H-bonding interactions are predicted to reflect the uncooperative characteristics, which is confirmed by high-precision ab initio calculations. These findings provide a direction for the comprehensive understanding of H-bonds.

## Main

The cooperative effect, as one of the most remarkable features of hydrogen bonds (H-bonds, referring to D-H···A), has always been the focus of researches[1-3]. However, how to obtain the mechanism of this effect is still of great challenge. Early, it was postulated qualitatively as that one H-bond would be strengthened, following by the formation of neighboring H-bond, which was further beneficial for the formation of more H-bonds[4,5]. Subsequently, the enhancement of H-bonds in the effect was described in terms of frequency shift of infrared spectrum[6] and interaction energy[7]. It was not until 1990's that the description of the cooperative effect transferred gradually to the structural changes, especially D-H and D~A distances[8]. So far, the mechanism of cooperative effect in water interaction systems has been empirically summarized as followed: with the decreasing of the intermolecular distance (D~A distance), the D-H bond length first increases and then decreases, while H···A bond length decreases[9,10]. In particular, although the cooperative effect in complex systems reflects many complications, such as facilitating the formation of large-size clusters[11], determining the properties of liquid water[12], ice[13] and DNA double helices[14], which are already well understood empirically at the atomic level. Not only that, the cooperative effect has been also confirmed to be consistent with the calculations based on first-principles density functional theory (DFT) and earlier parametric methods[15,16], thus exhibiting the fundamental significance. Nevertheless, in 2021, using a high-precision ab initio method, we reached the opposite conclusion in the study of water intermolecular interactions. As the D~A distance decreases, the D-H bond length always decreases without increasing, and the H···A bond length decreases. The phenomenon, different from the cooperative effect investigated previously, is called the uncooperative effect[17]. This draws considerable concern about whether the basic knowledge of H-bonding interactions has been sufficiently comprehensive over the past 60 years. Especially considering that both cooperative and uncooperative effects involve the essential understanding of H-bonding and even intermolecular interactions, it is urgent to form a clear grasp.

Actually, the structures and interactions of H-bonds may be two critical insights for understanding the cooperative and uncooperative effects. As far as previous studies of H-bonded structures are concerned, the H-bond formation is accompanied by the weakening and lengthening of D-H bond[18-20]. As for components of interactions, the H-bond traditionally tends to be highly electrostatic (i.e., dipole-dipole)[21]. However, several long-term but not sufficiently spotlighted views have suggested that H-bond is partly covalent and the corresponding covalency can be attributed to induced interaction[22-24]. Combining the current understandings of the structures and interactions, the cooperative effect of H-bond is considered to be related to polarization, charge transfer, electrostatics, covalency, etc[9,25-27]. These show that the complexity of cooperative and uncooperative effects may lie in the complexity of H-bond. More to the point, applying the quantum mechanical methods to solve problems at the atomic level, the intra- and intermolecular interactions associated with the complexity of H-bonds have been proven to correspond to different levels of electron correlation[28,29]. This implies that strict high-precision ab initio methods can be regarded as the benchmark. Even though previous studies have shown that cooperative effect is consistent with the results of DFT and empirical force methods[15,16], it is still necessary that the benchmark methods are applied to check, which should be taken seriously.

In this work, we used the well-accepted high-precision ab initio benchmark method, that is coupled-cluster singles and doubles with perturbative triple excitations CCSD(T), to explore abundant ubiquitous H-bonded systems. The calculations show that there is cooperative effect in elementary H-bonded systems with different D and A atoms (D≠A), while there is uncooperative effect in those with the same D and A (D=A). As importantly, we found that the two effects can be determined by the core-valence bifurcation (CVB) index and binding energy of the equilibrium H-bonded systems. Furthermore, this finding is supported by high-precision ab initio calculations of non-H-bonded interaction systems. Our work not only reveals the generality of cooperative and uncooperative effects in intermolecular interaction systems but also provides an important perspective for understanding intermolecular interactions in the future.

## Results and discussion

We first optimized different elementary H-bonded systems (see Figure 1). Those structures with different D and A atoms (D≠A) include F-H⋯$OH_2$, F-H⋯$OCH_2$, HO-H⋯$NH_3$, and F-H⋯$NH_3$, and others with the same D and A atom (D=A) include F-H⋯FH, HO-H⋯$OCH_2$, HO-H⋯$OH_2$ and $H_2$N-H⋯$NH_3$ (for details, see Supplementary Information (SI), Part 1).

The different D~A distances at the equilibrium position are taken as the zero point to further study the structures (Figures 1b and c) and energies (Figures 1b′ and c′) during compression. The calculations indicate that for the H-bonded systems of D≠A, with the decreasing of intermolecular distance, the D-H bond length first increases and then decreases, which reproduces the common cooperative effect[13,14]. Here, to make the results explicit, we mainly focus on the variation of D~A distances ranging between 0 and -0.24 Å, and their complete contraction process (from 0 to -0.51 Å) is shown in Parts 2 and 3 of the SI. However, in the H-bonded systems of D=A, the D-H bond length decreases without increasing. In contrast to previous cooperativity, this abnormal phenomenon is called the uncooperative effect[17]. At this point, as the D~A distance gradually decreases by 0.24 Å, the D-H bond length increases by a magnitude of $10^{-4}$ Å for the cooperative effect and deceases by a magnitude of $10^{-4}$ Å for the uncooperative effect. Additionally, the binding energies of the two effects increase monotonically. Of particular note, the results imply that the two effects are general in H-bonded systems. In the following, we will analyse their properties and look for reference laws. There are two questions have attracted our attention: Why are the two effects related to whether D and A are the same? Can we predict the properties of other intermolecular interaction systems?

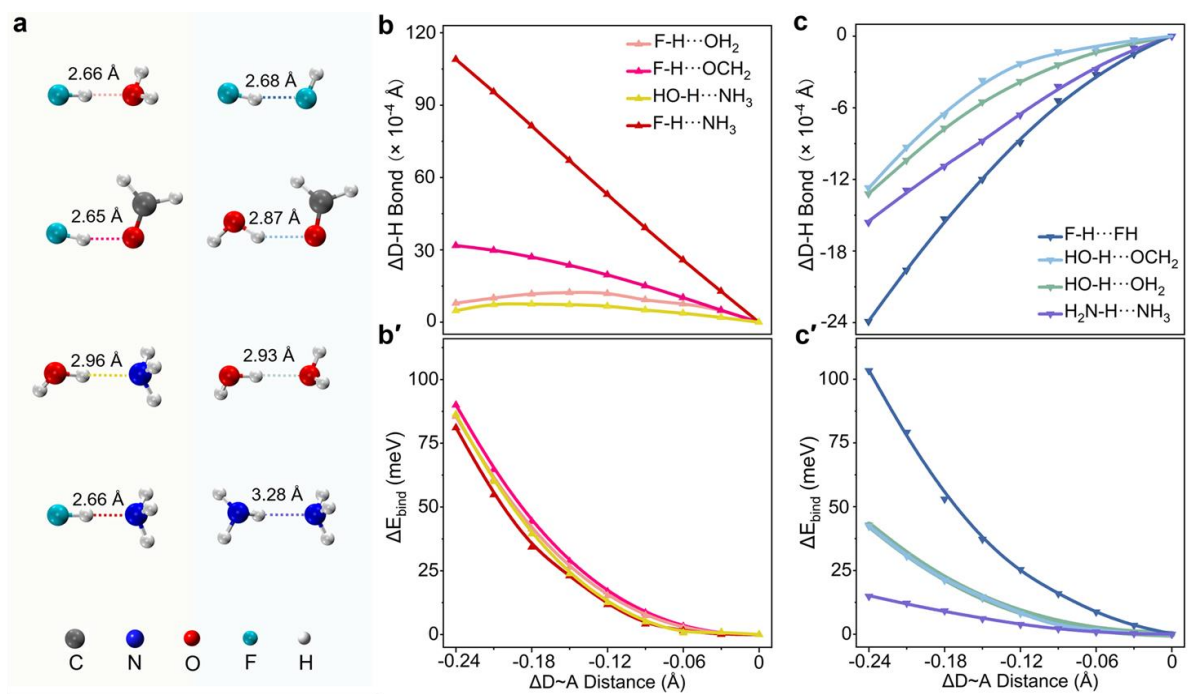

**Figure 1 | The structures and contraction properties of intermolecular H-bonded systems. a,** The structures of H-bonded systems. Corresponding D~A distances are displayed. **b** and **b′,** Changes in D-H bond lengths and binding energies as the D~A

distance decreases in H-bonded systems with D≠A. With the contraction of the D~A distance, the D-H bond length first increases and then decreases, as shown in Part 4 of the SI. Here, the D~A distances ranging between 0 and -0.24 Å are highlighted. **c** and **c',** Changes in D-H bond lengths and binding energies as the D~A distance decreases in H-bonded systems with D=A.

To explore these questions, the CVB index and binding energy of equilibrium structures were analysed. According to Table S49 and Figure 2a, compared to the D=A systems, D≠A systems have a more negative CVB index and higher released binding energies. Taking F-H⋯$NH_3$ as an example, it has the most negative CVB index and highest released binding energy, thereby reflecting the strongest H-bond among the studied structures. Specifically, the uncooperative H-bonded systems release lower binding energies (<0.25 eV) and have a greater CVB index (>0.022), which are remarkably distinct from cooperative systems. By incorporating all data points into the same figure, the horizontal coordinate axis X denotes the binding energy, and the vertical coordinate axis Y represents the CVB index. We found that these points can be fitted linearly as y = a·x+ b (see Figure 2c), where a=-0.19 and b=0.07, manifesting the inverse relationship between the CVB index and binding energy. It has been investigated that the linearity is by no accident. The reason is that the CVB index also directly contains information on the H-bond strength[30,31]. A smaller CVB index corresponds to more released binding energies, which also means a stronger H-bond[32]. The above results show that cooperative H-bonded systems (D≠A) perform stronger H-bonds, while uncooperative systems (D=A) have weaker H-bonds.

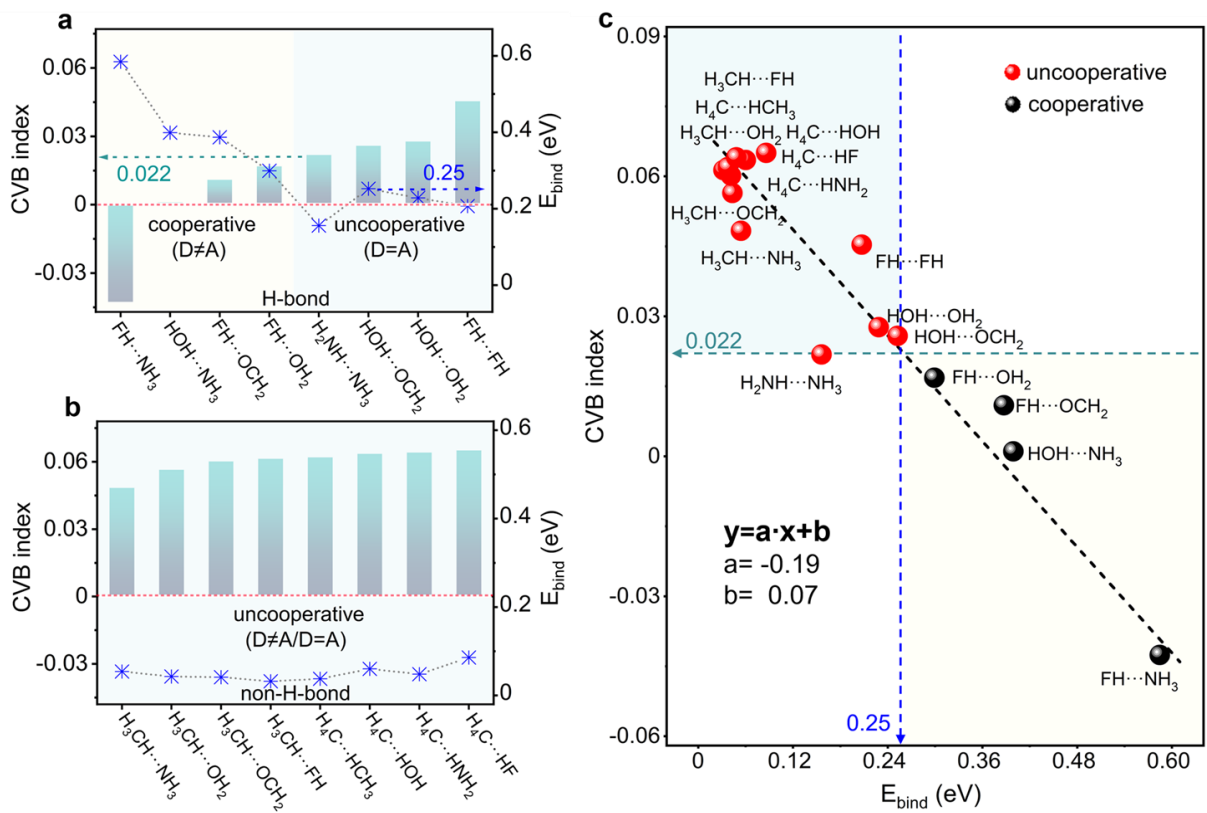

**Figure 2 | The core-valence bifurcation (CVB) index and binding energies of equilibrium structures. a,** The CVB index and binding energies of H-bonded systems. **b,** The CVB index and binding energies of non-H-bonded systems. Between them, the CVB index, corresponding to the bar, is marked on the left vertical axis, and the binding energy, corresponding to the scatter, is marked on the right vertical axis. The blue dotted line indicates that the binding energy of HO-H···OCH$_2$ is 0.25 eV, and the green dotted line denotes that the CVB index of H$_2$N-H···NH$_3$ is 0.022. **c,** Relationship between binding energy and CVB index and corresponding linear fitting formula. The coefficient a and constant b are -0.19 and 0.07 for our simulations, respectively. The values 0.25 eV and 0.022 denote the boundary of cooperative and uncooperative effects. The light yellow and blue areas in the figure represent cooperative and uncooperative effects, respectively.

Generally speaking, non-H-bonded intermolecular interactions release lower binding energies. The intermolecular interactions can be regarded as different levels of electron correlation[33,34]. It can be reasonably speculated that non-H-bonded systems may exhibit the uncooperative effect. For this reason, equilibrium non-H-bonded systems were investigated. In comparison with H-bonded systems, these typical non-H-bonded systems (see Figure 3b), including H$_3$C-H···FH, H$_3$C-H···OH$_2$, H$_3$C-H···OCH$_2$, H$_3$C-H···NH$_3$, F-H···CH$_4$, HO-H···CH$_4$, H$_2$N-H···CH$_4$ and H$_3$C-H···CH$_4$, have a greater CVB index and release lower binding energies, which indeed means weaker intermolecular interactions.

Next, the non-H-bonded systems are compressed to confirm whether they have uncooperative effects and whether they conform to our general judgement. Hence, the structures and energies of all these non-H-bonded systems during compression were studied. As shown in Figure 3, the calculations distinctly demonstrate that these non-H-bonded systems conform to the uncooperative effect. Specifically, as the D~A distance gradually decreases by 0.24 Å, the D-H bond length decreases by a magnitude of $10^{-4}$ Å. Comparing the D-H bond lengths in cooperative and uncooperative systems, the former increases by $10^{-4}$ Å magnitude, and the latter decreases by a magnitude of $10^{-4}$ Å. These results further reveal that two effects are related to the H-bond strength. Their intermolecular interaction properties can be predicted just by analysing the corresponding equilibrium structures. This finding provides a deeper viewpoint into intermolecular interactions, especially H-bonding interactions.

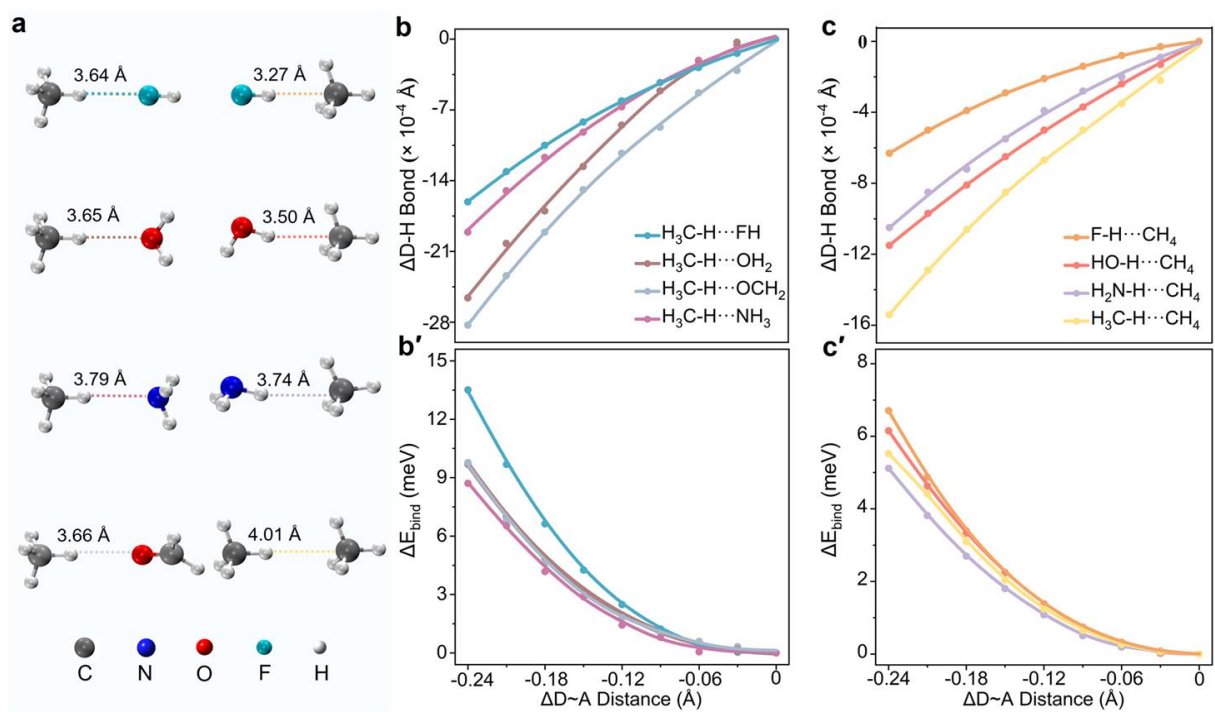

**Figure 3 | The structures and contraction properties of typical non-H-bonded systems. a,** The structures of non-H-bonded systems. Corresponding D~A distances are shown. **b** and **b',** Changes in the D-H bond lengths and the binding energies in the non-H-bonded systems with C as the donor. **c** and **c',** Changes in the D-H bond lengths and binding energies in the non-H-bonded systems with C as the acceptor.

Based on the above, as a direct characterization of the intermolecular interaction strength, the binding energy is the key to exploiting the cooperative and uncooperative effects. Therefore, as a supplement, using the energy decomposition analysis (EDA) method, we further analyzed the components of binding energy in all intermolecular interaction systems of this work. In EDA, the total interaction $E_{int}$ can be explicitly decomposed into four parts, namely, the exchange repulsion $E_{ex}$, electrostatic $E_{elec}$, induced $E_{ind}$ and dispersion $E_{disp}$ interactions, and the details are shown in Part 5 of the SI. The last three terms and their percentage contributions to the total attractive interaction $E_{attr}$ are displayed in Figures 4a-c and a'-c', respectively. For uncooperative systems, the four components all exhibit considerably smaller interaction energies than those for cooperative systems. The weaker interaction mirrors a weaker H-bond. In other words, the uncooperative systems more easily form and maintain dimers, which is in reasonable accordance with the aforementioned reference laws and previous experimental observations[35,36]. The interaction energy values of the two effects have apparent boundaries. However, the boundary disappears regarding the percentage of the $E_{elec}$. Meanwhile, the $E_{ex}$ of cooperative systems is greater, which violates cooperative attractive character[27]. Therefore, $E_{elec}$ and $E_{ex}$ may not be the real reason for the two effects. The $E_{ind}$ and $E_{disp}$ need to be further discussed.

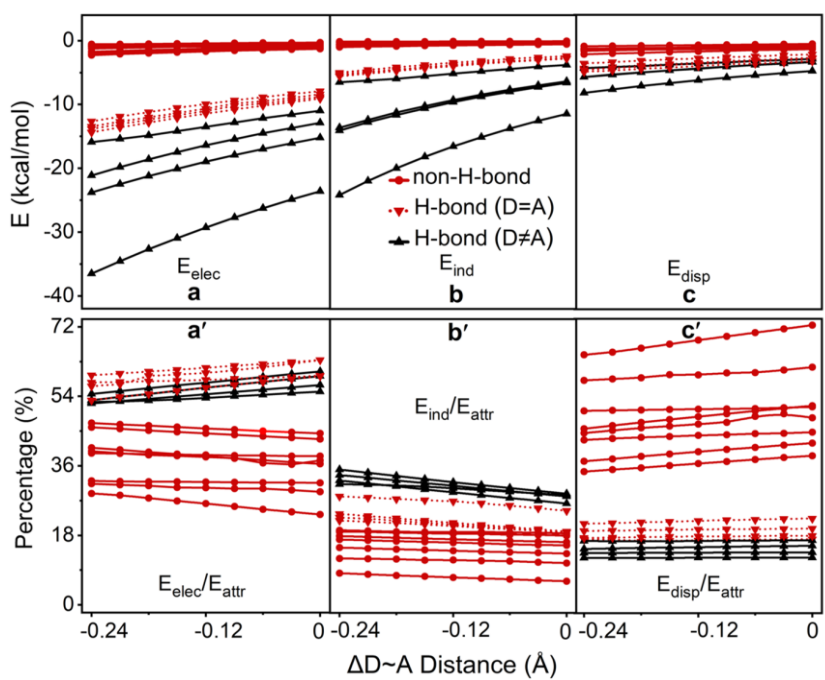

**Figure 4 | Energy decomposition analysis during the contraction of different systems. a, b and c,** Changes in the

electrostatic $E_{elec}$, induced $E_{ind}$ and dispersion $E_{disp}$ interaction energy as the D~A distance decreases, respectively. **a', b' and c',** Changes in the percentage contributions of $E_{elec}$, $E_{ind}$ and $E_{disp}$ to attractive interaction $E_{attr}$ as the D~A distance decreases, respectively. The black solid lines represent H-bonded systems with D≠A. The red dashed and solid lines represent H-bonded systems with D=A and non-H-bonded systems, respectively.

The $E_{ind}$ and $E_{disp}$ interaction terms were further analysed, as epitomized in Figure 4a'-b'. The cooperative systems show a greater $E_{ind}$ percentage than uncooperative systems. The general phenomena of $E_{ind}$ ensure that cooperative systems display more attractive interaction[27]. As a matter of fact, $E_{ind}$ accounts for charge transfer and polarization[37], and takes advantage in the stability of systems[38]. This can provide a possible explanation that the cooperative effect exists in H-bonded systems with D≠A, while the uncooperative effect is found in those H-bonded systems with D=A. Again, the SAPT method corrects high-order $E_{ind}$ and coupling between $E_{ind}$ and $E_{disp}$ to further ensure the reliability of our conclusions. Further, taking HO-H···OH$_2$ and F-H···OH$_2$ as examples, the induced interaction also plays a unique role during the reverse stretch and contraction of the D-H bond length relative to the equilibrium position as the D~A distance decreases, as shown in Part 6 of the SI. These findings lead us to conclude that the $E_{ind}$ may be the real reason of these two effects.

In summary, our work proves the generality of uncooperative and cooperative effects in intermolecular interactions, especially in elementary H-bonding interactions. The high-precision ab initio calculations show that the well-known cooperative effect is reproduced in H-bonded systems with different D and A atoms (D≠A), while uncooperative effect is found when D and A are the same (D=A). Further analysis suggests that the two effects can be quantitatively distinguished by the binding energy and CVB index. Accordingly, we predict that intermolecular non-H-bonded interaction systems can reflect uncooperative characteristics, and confirm this conclusion by high-precision ab initio calculations. These findings will provide an essential reference for the future study of intermolecular interactions such as H-bonding interactions.

## METHODS

To carry out this study, the geometries of the studied systems were fully optimized at the ab initio CCSD(T) level in conjunction with the aug-cc-pVDZ basis set using the MOLPRO 2012 program[39]. The CCSD(T) method, as a benchmark method for high-precision ab initio calculations, has great advantages in theoretical research[40-42]. It corrects the electronic correlation energy and contains multinomial configuration interactions of electron excitation. Therefore, this method can be applied in intermolecular interactions, especially H-bonding interactions. To clarify the essence of the two effects in equilibrium structures, the core-valence bifurcation (CVB) index[43] based on topological analysis of the electronic localization function (ELF)[44] using Multiwfn program[45] and binding energies between these two monomers were also computed at the same level.

The interaction energy, that is, binding energy, was decomposed by EDA method[46] with the Psi4 program[47] to get an insight for the essence of the cooperative and uncooperative effects. Furthermore, the method is based on Symmetry-Adapt Perturbation Theory (SAPT)[48,49]. Total interaction energy ($E_{int}$) between two monomers was decomposed into four interaction terms with clearly physical pictures: the exchange repulsion interaction energy ($E_{ex}$), electrostatic interaction energy ($E_{elec}$), induced interaction energy ($E_{ind}$) and dispersion interaction energy ($E_{disp}$). Therefore, the $E_{int}$ between two monomers can be defined as:

$$E_{int} = E_{ex} + E_{elect} + E_{ind} + E_{disp}$$

Additionally, $E_{ex}$ makes repulsive contributions, the $E_{elec}$, $E_{ind}$ and $E_{disp}$ make attractive contributions. $E_{attr}$ represents the total attractive interaction, which denotes the sum of $E_{elec}$, $E_{ind}$ and $E_{disp}$. Corresponding percentage values represent every interaction contribution to $E_{attr}$ (i.e., $E_{elec}$, $E_{ind}$ or $E_{disp}/E_{attr}$).

## Data availability

The data that support the findings of this study are available from the corresponding author upon reasonable request.

## Acknowledgements

This work was supported by 2020-JCJQ project（GFJQ2126-007）and the National Natural Science Foundation of China (grant numbers 11974136). Z. Wang also acknowledges the assistance of the High-Performance Computing Center of Jilin University and National Supercomputing Center in Shanghai.

## Author contributions

R. Liu performed the theoretical simulations. Z. Wang supervised the work. R. Liu, R. Wang, D. Li, Y. Zhu, X. Yang and Z. Wang analyzed the results. R. Liu and Z. Wang wrote the article.

## Competing interests

All authors declare no competing interests.